\documentclass[sigconf]{acmart}

\usepackage{multirow}
\usepackage{booktabs}
\usepackage{amsmath,amssymb,amsfonts}
\usepackage{algorithmic}
\usepackage{graphicx}

\usepackage{tikz}
\usepackage{pgfplots}

\AtBeginDocument{%
  \providecommand\BibTeX{{%
    \normalfont B\kern-0.5em{\scshape i\kern-0.25em b}\kern-0.8em\TeX}}}

\copyrightyear{2020}
\acmYear{2020}
\setcopyright{acmcopyright}
\acmConference[ICPC '20]{28th International Conference on Program Comprehension}{October 5--6, 2020}{Seoul, Republic of Korea}
\acmBooktitle{28th International Conference on Program Comprehension (ICPC '20), October 5--6, 2020, Seoul, Republic of Korea}
\acmPrice{15.00}
\acmDOI{10.1145/3387904.3389261}
\acmISBN{978-1-4503-7958-8/20/05}

\begin{document}

\title{A Self-Attentional Neural Architecture for Code Completion with Multi-Task Learning}


\author{Fang Liu$^{1,2}$, \ Ge Li$^{1,2 \dagger}$, \ Bolin Wei$^{1,2}$, \ Xin Xia$^3$, \  Zhiyi Fu$^{1,2}$, \ Zhi Jin$^{1,2 \dagger}$
} 
\thanks{$\dagger$ Corresponding author.}

\affiliation{
\institution{$^1$Key Laboratory of High Confidence Software Technologies (Peking University), Ministry of Education}
}

\affiliation{
\institution{$^2$Institute of Software, EECS, Peking University, Beijing, China}
}

\affiliation{
\institution{$^3$Faculty of Information Technology, Monash University, Australia}
}

\email{{liufang816, lige, ypfzy, zhijin}@pku.edu.cn, bolin.wbl@gmail.com, xin.xia@monash.edu}

\renewcommand{\shortauthors}{Liu et al.}

\begin{abstract}
Code completion, one of the most useful features in the Integrated Development Environments (IDEs), can accelerate software development by suggesting the libraries, APIs, and method names in real-time. Recent studies have shown that statistical language models can improve the performance of code completion tools through learning from large-scale software repositories. However, these models suffer from three major drawbacks: a) The hierarchical structural information of the programs is not fully utilized in the program's representation; b) In programs, the semantic relationships can be very long. Existing recurrent neural networks based language models are not sufficient to model the long-term dependency. c) Existing approaches perform a specific task in one model, which leads to the underuse of the information from related tasks. To address these challenges, in this paper, we propose a self-attentional neural architecture for code completion with multi-task learning. To utilize the hierarchical structural information of the programs, we present a novel method that considers the path from the predicting node to the root node. To capture the long-term dependency in the input programs, we adopt a self-attentional architecture based network as the base language model. To enable the knowledge sharing between related tasks, we creatively propose a Multi-Task Learning (MTL) framework to learn two related tasks in code completion jointly. Experiments on three real-world datasets demonstrate the effectiveness of our model when compared with state-of-the-art methods.
\end{abstract}






\maketitle

\begin{figure*}[t] 
\setlength{\abovecaptionskip}{0cm}
\centering\includegraphics[width=14cm]{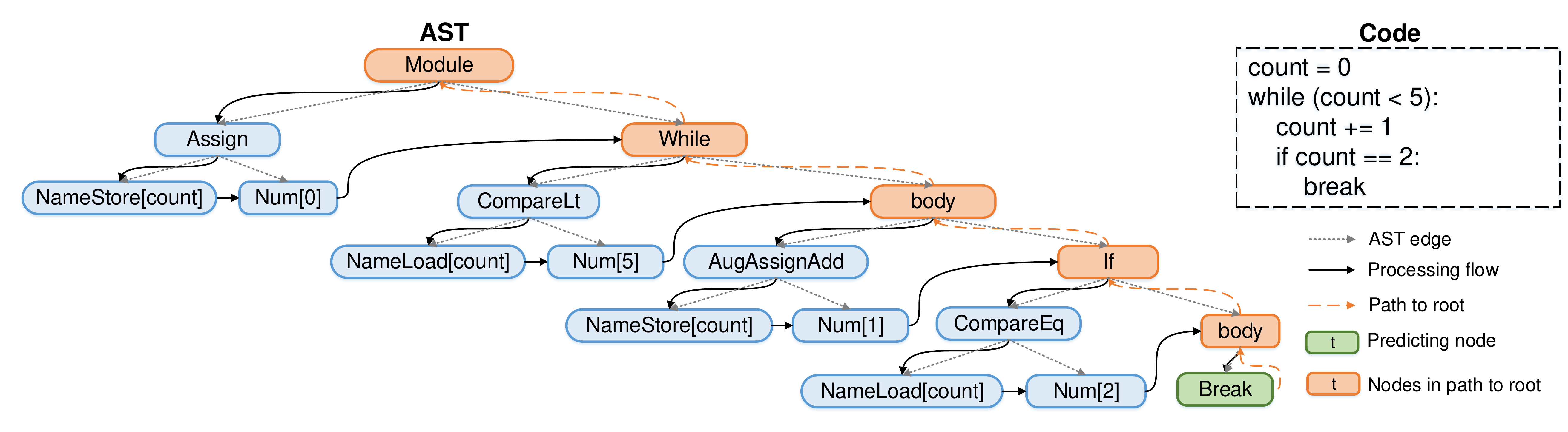} 
\caption{The AST of the given Python code snippet. Green node denotes the predicting node, i.e., \textit{Break}. Solid arrows indicate the nodes' processing order. Orange dotted arrows show the path from the predicting node to the root node.}
\label{Fig:AST}
\vspace{-0.3cm}
\end{figure*}

\section{Introduction}
As the complexity and scale of the software development continue to grow, code completion has become an essential feature of Integrated Development Environments (IDEs). It speeds up the process of software development by suggesting the next probable token based on existing code. However, traditional code completion tools rely on compile-time type information or heuristics rules to make recommendations \cite{liu2016neural, Li2018Code}, which is costly and could not capture human's programming patterns well. To alleviate this problem, code completion research started to focus on learning from large-scale codebases in recent years.

Based on the observation of source code's repeatability and predictability \cite{hindle2012naturalness}, statistical language models are generally used for code completion. N-gram is one of the most widely used language models \cite{hindle2012naturalness,tu2014localness,hellendoorn2017deep}. Most recently, as the success of deep learning, source code modeling techniques have turned to Recurrent Neural Network (RNN)-based models \cite{bhoopchand2016learning,Li2018Code}. In these models, a piece of source code is represented as a source code token sequence or an Abstract Syntactic Tree (AST) node sequence. Given a partial code sequence, the model computes the probability of the next token or AST node and recommends the one with the highest probability. However, these models are limited from three aspects: 

a) \textbf{The hierarchical structural information is not fully utilized in the program's representation}. Existing code completion models mainly fall into two major categories, i.e., token-based models and AST-based models. The token-based models \cite{bhoopchand2016learning,hellendoorn2017deep} sequentially tokenize programs into token sequences as the input of models. The syntax and structure of code are not explicitly considered, so this information is underused. To address this limitation, AST-based neural network models are proposed \cite{liu2016neural,Li2018Code}. In these models, programs are first parsed into ASTs. Then, ASTs are traversed to produce the node sequence as the representation of the programs. Although these models utilize ASTs in the program's representation, the hierarchical level of the AST nodes is ignored because the tree is traversed to flatten sequence. The tree's structural information is under-utilized. 

b)\textbf{ In programs, the semantic relationships might be very long}. For example, when the model suggests calling a function that has been defined many tokens before (e.g., 500 tokens). The parse tree of a program is typically much larger than that of a natural language sentence \cite{mou2016convolutional}. There are approximately 1730 nodes on average in JavaScript dataset of our experiment. In such a case, recent code completion work which builds LSTM-based language models \cite{bhoopchand2016learning,Li2018Code} cannot work on modeling the very long-term dependency in the source code well, since LSTM-based language models use 200 context words on average \cite{khandelwal2018sharp}. 

c)\textbf{ Current code completion models train a single model to perform a specific task}, e.g., predicting the next node's value in AST (i.e., predicting the next token of a program). In code completion, the node's type and value are two closely related attributes, where the type can serve as a constraint to the value, and vice versa. However, this correlation is not well considered in existing code completion models. Li et al. \cite{Li2018Code} built two models to predict node's type and value separately, and they treated these two tasks independently. We argue that the relationship among related tasks could provide effective constraints for each task's learning process, and knowledge obtained from one task might help the other task. Therefore, these tasks should not be learned separately. 

In this paper, we propose a self-attentional neural architecture for code completion with Multi-Task Learning (MTL) \cite{caruana1997multitask} to address the aforementioned three limitations. To bridge the gap between the sequential node sequences and the hierarchical structure of ASTs, we extract the path from the predicting node to the root node, which indicates the hierarchical level of the predicting node. Previous studies did not consider the hierarchical level into their code completion models. Then we model the path information into the representation of the contextual program. To capture the long-term dependency in the input programs, we apply the Transformer-XL network \cite{transformer-xl} as our base model. To enable the knowledge sharing between related tasks, we adopt MTL to learn two tasks together, i.e., predicting the next node's type and value, which are two main closely related tasks in code completion. MTL can help the model focus its attention on the features that actually matter as other tasks provide additional evidence for the relevance or irrelevance of those features, thus can further improve the model's performance.

To evaluate the performance of our proposed model, we conduct experiments on three real-world datasets, including Python, Java, and JavaScript, and compare our model with two state-of-the-art models: Nested N-gram model \cite{hellendoorn2017deep} and Pointer Mixture Network \cite{Li2018Code}. For the next node's type prediction, our model achieves the accuracy of 87\%, 82\%, and 91\% on these three datasets respectively, which improves Nested N-gram model by 51\%, 40\%, and 72\%, and improves Pointer Mixture Network by 33\%, 24\%, and 24\%, in terms of \textit{normalized improvement in accuracy}. For the next node's value prediction, our model achieves the accuracy of 73\%, 73\%, and 83\% on three datasets, which improves Pointer Mixture Network by 16\%, 15\%, and 13\%, in terms of \textit{normalized improvement in accuracy}. Statistical testing shows that the improvements over the baseline methods are statistically significant, and the effect sizes are non-negligible. The main contributions of this paper are summarized as follows:
\begin{itemize}
\item We propose a novel method that models the hierarchical structural information into the program's representation.
\item We invent a new multi-task learning model for code completion, which enables knowledge sharing between related tasks. To the best of our knowledge, it is the first time that a multi-task learning model is proposed to solve the code completion problem.
\item We introduce the Transformer-XL network as the language model to capture the very long-range dependencies for code completion.
\item We evaluate our proposed model on three real-world datasets. Experimental results show that our model achieves the best performance compared with the state-of-the-art models. 
\end{itemize} 

\noindent\textbf{Paper Organization} ~ The remainder of this paper is organized as follows. We give a motivating example in Section \ref{motivation} and provide background knowledge on statistical language model and multi-task learning in Section \ref{background}. Then we introduce our proposed model in Section \ref{approach}. Section \ref{exp} presents experimental results. Section \ref{Discussion} analyzes the efficiency and quality of our model and discusses threats to validity. Section \ref{related work} highlights the related work. Finally, we conclude our study and mention future work in Section \ref{conclusion}.

\section{Motivating Example}\label{motivation}
Figure \ref{Fig:AST} shows an AST of a Python code snippet. Each node in the AST contains a \textit{Type} attribute, and the leaf nodes also contain an optional \textit{Value} attribute. We use ``Type[Value]" to represent each node. To make full use of the structural information of the AST in the program's representation, we take the path from the predicting node to the root node into consideration, which indicates the hierarchical level of the predicting node. For example, in Figure \ref{Fig:AST}, when predicting the node \emph{Break}, the contextual sequence contains all the nodes in the tree except \emph{Break} if the tree is flattened in the in-order depth-first traversal \cite{liu2016neural,Li2018Code} (marked by solid black arrows in the figure). The hierarchical level of the predicting node is ignored. If the path from the predicting node \emph{Break} to root node (marked by orange arrows in the figure) is introduced into the program's representation explicitly, i.e., \emph{\{body, If, body, While, Module\}}, the structural level of the predicting node can be utilized. The model will realize that the predicting node is in the \emph{If} statement which is nested in the \emph{While} statement. This information would be helpful in code completion.

For the model's learning mechanism, training different models to predict node's type and value separately ignores the correlations between these tasks. These two tasks are closely related. For example, in Figure \ref{Fig:AST}, when the model is going to predict the node \textit{Num[0]}, the node's type ``Num" conveys the message that the node's value is a number. The model will probably predict a number as the node's value. Likewise, if the model knows the node's value is a number, the model will probably predict ``Num" as its type. Similarly, when predicting the node \textit{NameLoad[count]}, the type ``NameLoad" implies the information of variable accessing, which helps the model to predict a variable that has been defined as the node's value.
Based on the above analysis, we believe that related tasks should be learned jointly. In such a way, the model could learn their common features and achieve better performance.  

\begin{figure*}[ht] 
\setlength{\abovecaptionskip}{0.2cm} 
\centering\includegraphics[width=15cm]{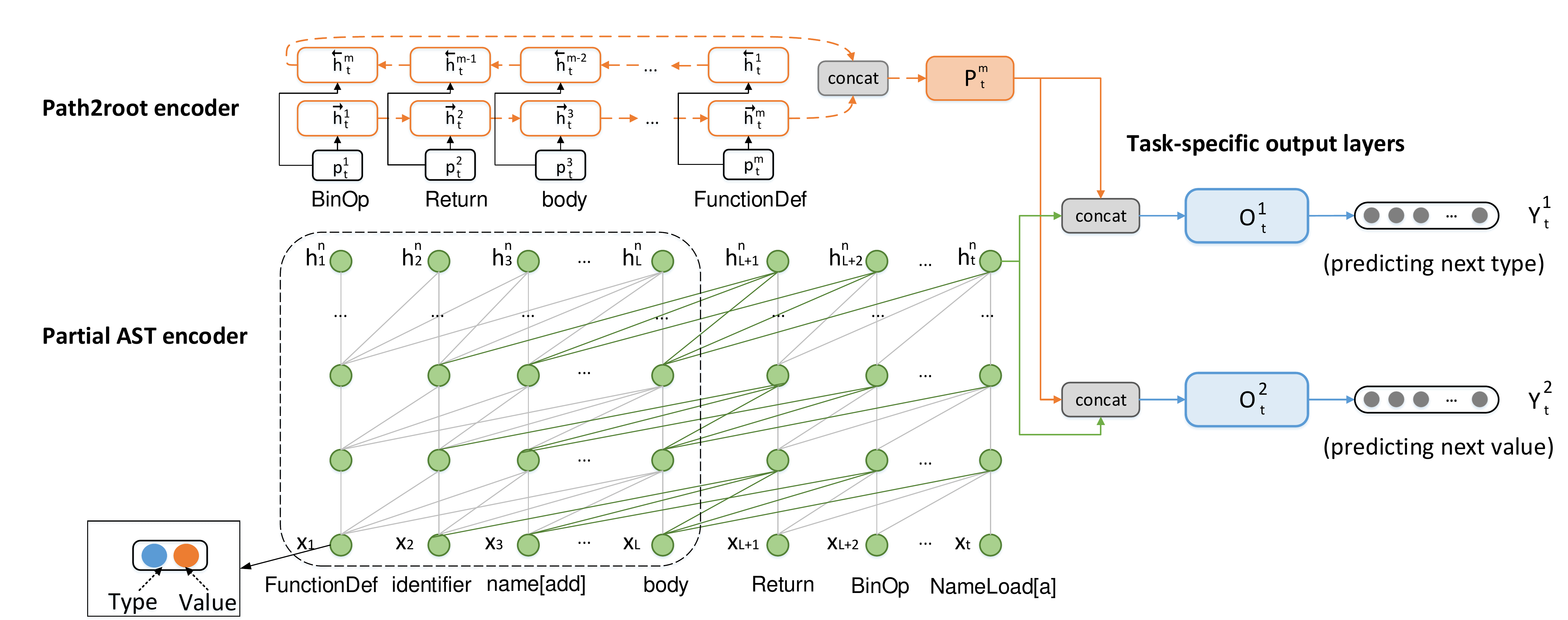} 
\caption{The architecture of our model, including partial AST encoder, path2root encoder and task-specific output layers.}
\label{Fig:model_Arch}
\vspace{-0.3cm}
\end{figure*}

\section{Background}\label{background}
In this section, we present the background knowledge which will be used in this paper, including the statistical language model and multi-task learning. 

\noindent\textbf{Statistical Language Model} ~ Statistical language models capture the statistical patterns in languages by assigning occurrence probabilities to a sequence of words in a particular sequence. Programming languages are kind of languages that contain predictable statistical properties \cite{hindle2012naturalness}, which can be modeled by statistical language models. Given a token sequence S = $s_{1},s_{2},...,s_{t}$, the probability of the sequence is computed as: 
\begin{equation}
p(S)=p(s_1)p(s_{2}|s_{1})p(s_{3}|s_{1}s_{2}),...,p(s_{t}|s_{1}s_{2},...,s_{t-1})
\end{equation}
The probabilities are hard to estimate when the number of the context tokens $s_{1},s_{2},...,s_{t-1}$ is tremendous. The N-gram model based on the Markov assumption is proposed to address this challenge. In the N-gram model, the probability of a token is dependent only on the $n-1$ most recent tokens:
\begin{equation}
p(s_{t}|s_{1},s_{2},...,s_{t-1})=p(s_{t}|s_{t-n+1},...,s_{t-1})
\end{equation}
N-gram based models have been generally applied to code completion \cite{hindle2012naturalness,tu2014localness,hellendoorn2017deep}. These models have been proved to capture the repetitive regularities in the source code effectively. In recent years, deep recurrent neural networks have shown great performance on modeling programming languages \cite{liu2016neural,bhoopchand2016learning,Li2018Code}. By using recurrent connections, information can cycle inside these networks for a long time, which loosens the fixed context size and can capture longer dependencies than the N-gram model. LSTM \cite{hochreiter1997long} and GRU \cite{cho2014properties} are two common variants of RNN, which ease the vanishing gradient problem in RNN by employing powerful gate mechanisms to remember and forget information about the context selectively. 

However, the introduction of gating in LSTMs might not be sufficient to address the gradient vanishing and explosion issue fully. Empirically, previous work has found that LSTM language models use 200 context words on average \cite{khandelwal2018sharp}, indicating room for further improvement. To ease this issue, attention mechanisms \cite{bahdanau2014neural,vaswani2017attention}, which add direct connections between long-distance word pairs, are proposed. For example, the Transformer \cite{vaswani2017attention} is an architecture based solely on attention mechanism. It uses a multi-headed self-attention mechanism to replace the recurrent layers to reduce sequential computation and capture longer-range dependency. However, the Transformer networks are limited by a fixed-length context in the setting of language modeling. To address this issue, Transformer-XL \cite{transformer-xl} is proposed by introducing the notion of recurrence into the deep self-attention network. Thus it enables the Transformer networks to model the very long-term dependency. In our model, we adopt Transformer-XL as the language model for the purpose of capturing the long-term dependency in programs.

\noindent\textbf{Multi-task Learning} ~ Multi-task learning is an approach for knowledge transfer across related tasks. It improves generalization by leveraging the domain-specific information contained in the training signals of related tasks \cite{caruana1997multitask}. It acts as a regularizer by introducing an inductive bias. As such, it reduces the risk of over-fitting \cite{ruder2017overview}.
There are two most commonly used ways to perform multi-task learning in deep neural networks: hard or soft parameter sharing of hidden layers. In soft parameter sharing, each task has its own hidden layers and output layer. To ensure the parameters of each task to be similar, the distance between the parameters of each task is regularized. Hard parameter sharing is the most commonly used way, where the hidden layers are shared among all tasks, and the output layers are task-specific. The shared hidden layers can capture the common features among all the tasks. Furthermore, by preferring the representation that all tasks prefer, the risk of over-fitting is reduced, and the model can be more general to new tasks in the future. To the best of our knowledge, MTL has not been applied to modeling source code. In this paper, we invent a novel MTL model to improve the performance of code completion.

\section{Proposed Model}\label{approach}
In this section, we first present an overview of the network architecture of our proposed model. Then we introduce each component of the model in detail.

\subsection{Overall Architecture}
Figure \ref{Fig:model_Arch} shows the architecture of our proposed model. At every point in the code (AST), our model gives a list of possible next nodes along with their probabilities that are estimated from the training corpus. We adopt Transformer-XL based language model as the partial AST encoder, which enables the Transformer network \cite{vaswani2017attention} to model very long-term dependency in the AST node sequence by introducing the recurrence into the deep self-attention network. We design a path2root encoder to capture the hierarchical information of the predicting node. Then we combine the output of the partial AST encoder and the path2root encoder together and use it to make predictions on the next node's type and value. MTL is adopted to learn these two tasks jointly. We argue that there exist some common features between these two tasks, and these features can be learned simultaneously. Thus, we employ the hard parameter sharing in our MTL framework, where the partial AST encoder and the path2root encoder are shared between tasks, and the task-specific output layers are used to produce task-specific outputs.

\subsection{Program Representation}
\begin{figure}[t] 
\setlength{\abovecaptionskip}{0cm}
\centering\includegraphics[width=9cm]{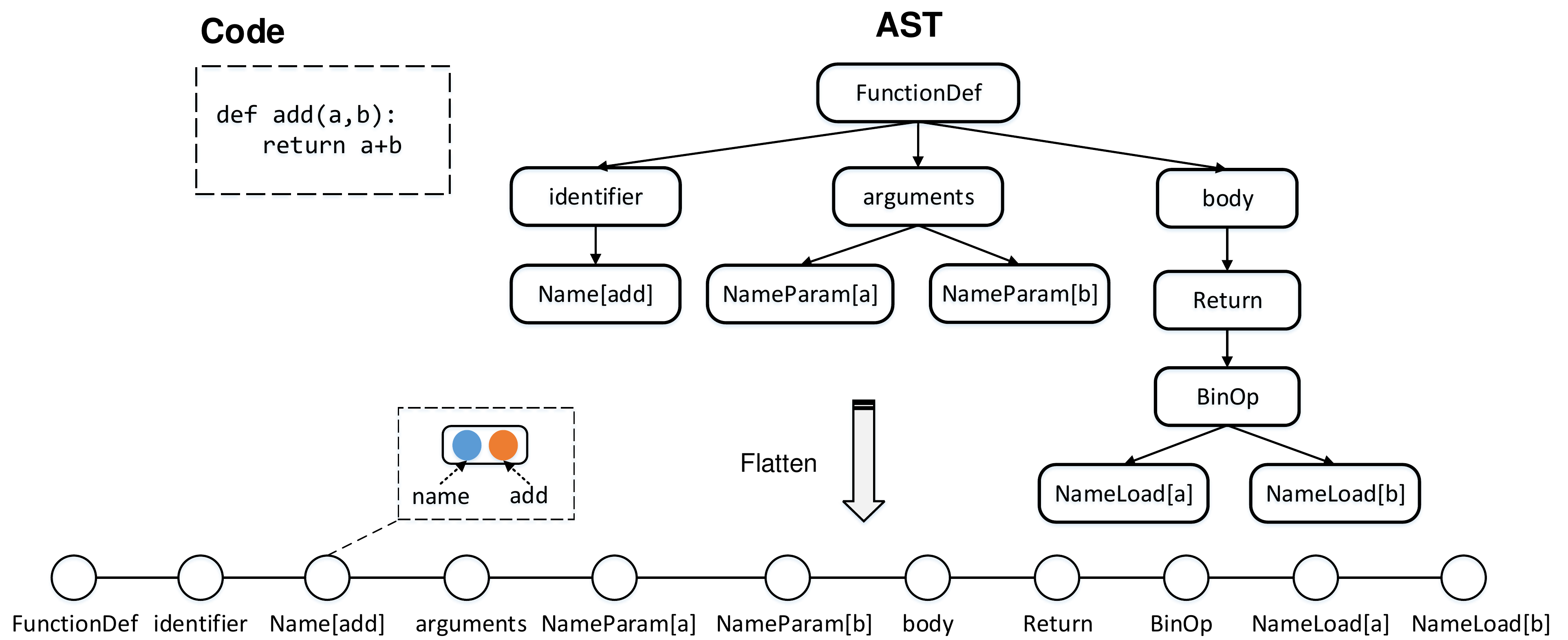}
\caption{Program representation.}
\vspace{-0.3cm}
\label{Fig:flatten}
\end{figure}

The programming language has an unambiguous context-free grammar, where each program can be parsed into a unique AST. ASTs are widely used for processing programs to extract the syntax and structure of programs \cite{mou2016convolutional,raychev2016probabilistic,Li2018Code}. We use ASTs to represent programs in our model and traverse them to node sequences. As shown in Figure \ref{Fig:flatten}, we use ``Type[value]" to represent each node. For non-leaf nodes that do not have the value attribute, we use a special symbol ``EMPTY'' to represent their value. We first flatten each AST in in-order depth-first traversal to produce a sequence of nodes. Then we represent both the \textit{Type} and \textit{Value} as real-valued vectors, and concatenate them as the final representation of the nodes $x_i = [T_i;V_i]$, where $T_i$ is the type vector, $V_i$ is the value vector, and ``;'' denotes the concatenation operation.

\subsection{Partial AST Encoder}
In our training and test datasets, the programs are represented as node sequences. The completion happens at every point in the node sequence, and the nodes before the point form as the contextual partial AST. \footnote{In practice, we can use existing tools such as jdt to parse the incomplete programs into incomplete ASTs by replacing the problematic nodes with some placeholders} We adopt the Transformer-XL network \cite{transformer-xl} to encode the partial AST, which captures the long-range dependencies in the sequence. In the vanilla Transformer language model, the length of the context is fixed. To address the limitations of using a fixed-length context, Transformer-XL is proposed to introduce a recurrence mechanism to the Transformer architecture. In Transformer-XL architecture, the hidden states of each new input segment are obtained by reusing that of the previous segments, instead of computing from scratch. In this way, the recurrent connection is created, and the reused hidden states can serve as memories for the current segment, which enables the information to propagate through the recurrent connections. Thus the model can capture very long-term dependency.

Formally, let $s_\tau=[x_{\tau,1},x_{\tau,2}, ..., x_{\tau,L}]$ and $s_{\tau+1}=[x_{\tau + 1,1},x_{\tau + 1,2}, \\ ..., x_{\tau + 1,L}]$ represent two consecutive segments of length $L$. For the $\tau$-th segment $s_\tau$, the $n$-th layer hidden state sequence is denoted as $h_\tau ^n \in \mathbb{R}^{L\times H}$, where $H$ is the dimension of the hidden units. The $n$-th layer hidden state for segment $s_\tau$ is computed as:

\begin{equation}
\begin{split}
& \widetilde{h}_{\tau+1}^{n-1}  = [SG(h_{\tau}^{n-1}) \circ h_{\tau+1}^{n-1}] \\
& q_{\tau+1}^{n}, k_{\tau+1}^{n}, v_{\tau+1}^{n} = h_{\tau+1}^{n-1}W_q^T, \widetilde{h}_{\tau+1}^{n-1}W_k^T, \widetilde{h}_{\tau+1}^{n-1}W_v^T \\
& h_{\tau+1}^n = \text{Transformer-Layer} (q_{\tau+1}^{n}, k_{\tau+1}^{n}, v_{\tau+1}^{n}) 
\end{split}
\end{equation}

where $SG(\cdot)$ stands for stop-gradient, that is, we don’t calculate gradients for the $\tau$-th segment. The notation $[h_u \circ h_v]$ indicates the concatenation of two hidden sequences along the length dimension, and $W_.^T$ denotes model parameters. Compared to the standard Transformer, the critical difference lies in that the key $k_{\tau+1}^{n}$ and value $v_{\tau+1}^{n}$ are conditioned on the extended context $\widetilde{h}_{\tau+1}^{n-1}$ and hence $h_{\tau+1}^{n-1}$ cached from the previous segment. The Transformer-layer consists of multi-head self-attention mechanism and a position-wise fully connected feed-forward network. Besides, to keep the positional information coherent when we reuse the states, relative positional embedding is adopted, and the detailed computation procedure can be found in \citet{transformer-xl}.

\subsection{Path2root Encoder}
To model the hierarchical structural information of the predicting node, we extract the path from the predicting node to the root node, i.e., {$p_t^1, p_t^2, ..., p_t^m$}, where $m$ is the length of the path, $p_t^i$ is the type of the $i$-th node in the path at time step $t$. \footnote{The nodes in the path are non-leaf nodes, and they do not have the value attribute. Thus, we use the node's type as the representation for the nodes in the path.} Taking the AST in Figure \ref{Fig:flatten} as an example, when predicting the last node \textit{NameLoad[b]}, the path from it to the root node contains the nodes \emph{\{BinOp, Return, body, FunctionDef\}}. As shown in Figure \ref{Fig:model_Arch}, we design a bidirectional-LSTM \cite{schuster1997bidirectional} based path2root encoder, which encodes the nodes in the path to produce a path vector. The hidden states for both directions of the bi-LSTM are computed as follows: 
\begin{equation}
\begin{split}
& \overrightarrow{h_t^i} = \overrightarrow{LSTM}(p_t^i, \overrightarrow{h_t^{i-1}}) \\
& \overleftarrow{h_t^i} = \overleftarrow{LSTM}(p_t^i, \overleftarrow{h_t^{i-1}})  \\
\end{split}
\end{equation}
$\overrightarrow{h_t^m}$ and $\overleftarrow{h_t^m}$ contain the path's forward information and backward information. We concatenate $\overrightarrow{h_t^m}$ and $\overleftarrow{h_t^m}$ to obtain the final path vector $P_t$ for each time step, i.e., $P_t = [\overrightarrow{h_t^{m}};\overleftarrow{h_t^{m}}]$. In this way, we can reduce the chance that the model might forget the information of the top nodes or the bottom nodes when the path is long.  

\subsection{Task-specific Output Layers}
\noindent\textbf{Tasks.} ~ Given a sequence of AST nodes, the code completion model is adopted to predict the next node, including node's type and value. These two attributes are closely related and interacted. Therefore, in our model, we adopt MTL to learn these two tasks together.

\noindent\textbf{Output Layers.} ~ 
In our model, we adopt task-specific output layers to produce task-specific outputs. The output of the partial AST encoder $h_t^n$ and path vector $P_t$ are concatenated to compute the task-specific output vector $O_{t}^{k}$. \textit{Sotfmax} function can takes as input a vector of $N$ real numbers, and normalizes it into a probability distribution consisting of $N$ probabilities proportional to the exponentials of the input numbers. We use the \textit{softmax} function to produce the probability distribution of the outputs $Y_{t}^{k}$:
\begin{equation}\label{formula:y}
\begin{split}
O_{t}^{k} & = \tanh(W^o(h_t^n;P_t)) \\
Y_{t}^{k} & = \text{softmax}(W^y O_{t}^{k} +b^y) 
\end{split}
\end{equation}
where $W^o\in\mathbb{R}^{H \times (H+H_p)}, W^y\in\mathbb{R}^{V \times H},  b^y\in\mathbb{R}^{V} $ are trainable parameters. $V$ is the vocabulary size, $H_p$ is the hidden size of the path2root encoder, and ``;'' denotes the concatenation operation. 

\subsection{Training}
To learn the related tasks jointly, we adopt a weighted sum over the task-specific losses as the final loss:
\begin{equation}
loss = \sum_{k=1}^{N} \alpha_k \times loss_{k} 
\end{equation}
where N is the number of tasks. $\alpha_k$ is the weight of the loss for the $k$-th task, and $\alpha_k\geq0, \ \sum_{k=1}^{N} \alpha_k = 1$. In this paper, by default, we set the weights for the two tasks as 0.5 and 0.5, respectively. The effect of different weight settings will be discussed in Section \ref{Discussion}. 

\section{Experiments and Analysis}\label{exp}
In this section, we present the experiments and analysis. Firstly, we introduce the datasets and the experimental setup. Then we propose the two research questions and conduct experiments to answer them.

\subsection{Dataset and Metrics}
We evaluate our model on three datasets: Python, Java, and JavaScript. Python and JavaScript datasets are collected from GitHub repositories by removing duplicate files, removing project forks, keeping only programs that parse and have at most 30,000 nodes in the AST, and they are publicly available.\footnote{in http://plml.ethz.ch} Each dataset contains 100,000 training programs and 50,000 test programs. Both source code files and their corresponding ASTs are provided. These two datasets have been used in \citet{Li2018Code} and \citet{raychev2016probabilistic}. Java dataset comes from \citet{hu2018summarizing}, where the programs are also collected from Github. We randomly sample 100,000 Java programs for training and 50,000 for test. We use javalang \footnote{https://github.com/c2nes/javalang} to parse the programs into ASTs, and we make it public available.\footnote{https://drive.google.com/open?id=1xxnYAu8L5i6TpNpMNDWxSNsOs3XYxS6T} For all the datasets, each program is represented in its AST format, and the AST is serialized in in-order depth-first traversal to produce the AST node sequence. Then we generate queries used for training and test, one per AST node, by removing the node and all the nodes to the right from the sequence and then attempting to predict the node. The number of type attributes and value attributes of AST nodes, the queries of the programs, and the average length of the AST nodes in programs are shown in Table \ref{tab:datasets}.

\begin{table}[t]
\centering
\setlength{\abovecaptionskip}{0cm} 
\caption{Detailed information of datasets.}
\begin{tabular}{lccc}  
\toprule
  ~ & Python & Java & JavaScript \\
\midrule
\# of Type & 330 & 175 & 95 \\
\# of Value & $3.4 \times 10^6$ & $2.1 \times 10^6$ & $2.6 \times 10^6$ \\
\# of Training Queries & $6.2 \times 10^7$ & $2.6 \times 10^7$ & $10.7 \times 10^7$  \\
\# of Test Queries & $3.0 \times 10^7$ & $1.3 \times 10^7$ & $5.3 \times 10^7$\\
Avg. nodes in AST & 623 & 266 & 1730 \\
\bottomrule
\end{tabular}
\vspace{-0.3cm}
\label{tab:datasets}
\end{table}

\begin{table*}
  \begin{center}
  \setlength{\abovecaptionskip}{0cm} 
  \caption{Accuracy comparison of state-of-the-art approaches and our proposed model. The numbers in the bracket following the results of the baseline models show the \textit{normalized improvement accuracy} of our model over the baselines.}
    \begin{tabular}{lcccccc} 
      \toprule
      ~ & \multicolumn{2}{c}{Python} & \multicolumn{2}{c}{Java} & \multicolumn{2}{c}{JavaScript}\\
      ~ & Type & Value & Type & Value & Type & Value \\
      \midrule
      Nested Cache N-gram &  73.2\% (51.2\%) & - & 69.3\% (40.4\%) & - &  69.5\% (71.5\%) & - \\
      Pointer Mixture Network & 80.6\% (32.5\%) & 70.1\% (16.4\%) & 75.9\% (24.1\%) & 70.7\% (14.7\%) & 88.6\% (23.7\%) & 81.0\% (12.5\%)  \\
      Our Model & \textbf{86.9\%} & \textbf{73.2\%}  & \textbf{81.7\%} & \textbf{73.1\%} & \textbf{91.3\%} & \textbf{82.5\%} \\
      \bottomrule
    \end{tabular}
    \label{tab:results}
    \vspace{-0.3cm}
  \end{center}
\end{table*}

We use \textit{accuracy} to evaluate the performance of our model. In the code completion task, the model provides an ordered list of suggestions for each node's type or value in the source code file given the context. We compute the top-1 accuracy, i.e., the fraction of times the correct suggestion appears in the first of the predicted list. Directly comparing accuracies by the difference or direct proportion may lead to inflated results (>100\% improvement). Therefore, we also use \textit{normalized improvement in accuracy} (\textit{Imp. Accuracy})  \cite{costa2016tipmerge} to measure the ``the room for improvement'':

\begin{equation}
Imp. \ Accuracy = \left\{
\begin{aligned}
&\frac{Acc_x - Acc_y}{Acc_{ub}-Acc_y}, \text{if } Acc_x > Acc_y \\
&\frac{Acc_x - Acc_y}{Acc_y}, \text{otherwise}
\end{aligned}
\right.
\end{equation}
where $Acc_x$ represents the accuracy obtained by model $x$, $Acc_y$ represents the accuracy obtained by model $y$, and $Acc_{ub}$ represents the upper bound of the accuracy \footnote{For the next node's type prediction, the upper bound of the accuracy is 100\%. For the next node's value prediction, since the UNK targets are treated as wrong predictions, the upper bound of the accuracy is less than 100\%, which depends on the OoV rate of the dataset.}. Thus, this metric can measure the room for improvement of model $x$ over model $y$.

\subsection{Experimental Setup}
To make a fair comparison with \citet{Li2018Code}, we use the same parameters proposed in their paper, including embedding size, hidden size of the AST encoder, vocabulary size, etc. The embedding sizes for type and value are 300 and 1,200, respectively. Hence, the size of the AST node vector is $300 + 1200 = 1500$. As shown in Table \ref{tab:datasets}, the number of the value attribute is large. Followed by \citet{Li2018Code}, we choose the 50,000 most frequent values to build value's vocabulary for all the three datasets. For those values outside the vocabulary, we use \emph{UNK} (unknown values) to represent them. The \emph{UNK} rate for Python, Java, and JavaScript are 11\%, 13\%, and 7\%, respectively. All the types are used to build type's vocabulary.

For the partial AST encoder, we use a 6-layer Transformer-XL network \cite{transformer-xl}. We employ $h = 6$ parallel heads, and the dimension of each head $d_{head}$ is set to 64. We set the segment length to 50, which is the same as the LSTM's unrolling length (the length of the input sequence) in \citet{Li2018Code}. The dimensionality of the hidden unit is $H$ = 1500. Through the recurrent mechanism, we can cache previous segments and reuse them as the extra context when processing the current segment. Considering the GPU memory and training time, we set the length of cached hidden states $M$ to 256. In our experiment, as we increase $M$, the accuracy also increases. When $M$ is increased to 1024, the accuracy stops increasing, which demonstrates that our model can use up to about 1024 context tokens. For the LSTM-based model, the accuracy stops increasing when the unrolling length increases to 256, which demonstrates that LSTM language models can only use less than 256 contextual tokens in this experiment, which is consistent with the findings in \cite{khandelwal2018sharp}. 

The dimension of the feed-forward layer in the Transformer is set to 1024. For the path2root encoder, we employ a single layer bidirectional-LSTM. In our model, we set the length of the path to $m$. For the nodes whose length is over $m$, we preserve $m$ nodes in the path from the predicting node to the root. For the nodes whose length is less than $m$, we pad the path to the length of $m$. Considering the trade-off between time cost and performance, we set the length of path $m$ to 5 and the hidden size of path2root encoder and path vector size to 300, which can offer a considerable improvement and would not increase much time cost. 

To train the model, we employ the cross-entropy loss and Adam optimizer \cite{KingmaB14}. In both the training and test process, the predictions of the \textit{UNK} targets are treated as wrong predictions as in \citet{Li2018Code}. Each experiment is run for three times, and the average result is reported. The hyper-parameters are selected on the validation set, that is, we choose the hyper-parameters settings associated with the best validation performance. We implement our model using Tensorflow \cite{abadi2016tensorflow} and run our experiments on a Linux server with the NVIDIA GTX TITAN Xp GPU with 12 GB memory.

\begin{table*}
  \begin{center}
  \setlength{\abovecaptionskip}{0cm}
  \caption{Effectiveness of each component in our proposed model.}
  \setlength{\tabcolsep}{4mm}{
    \begin{tabular}{lcccccc}
      \toprule
      ~ & \multicolumn{2}{c}{Python} & \multicolumn{2}{c}{Java} & \multicolumn{2}{c}{JavaScript}\\
      ~ & Type & Value & Type & Value & Type & Value \\
      \midrule
      Full model & 86.9\% & 73.2\% & 81.7\% & 73.1\%  &  91.3\% &  82.5\% \\
      - MTL & 84.2\% & 71.8\% & 79.7\% & 71.6\% & 89.5\% & 80.8\% \\
      - Path2root Encoder & 84.8\% & 72.2\% & 80.1\% & 72.4\% & 90.6\% & 81.6\% \\
      - Recurrence & 80.4\% & 67.6\% & 76.1\% & 67.6\% & 85.8\% & 77.9\% \\
      \midrule
      vanilla Transformer-XL & 82.3\% & 69.8\% & 78.0\% & 70.6\% & 88.5\% & 80.1\% \\
      \bottomrule
    \end{tabular}
    }
    \label{tab:components}
    \vspace{-0.3cm}
  \end{center}
\end{table*}

\subsection{Research Questions and Results}
To evaluate our proposed approach, in this section, we conduct experiments to investigate the following research questions:

\textbf{\textit{RQ1: How does our proposed approach perform in code completion when compared with state-of-the-art models?}} ~ To answer this research question, we compare our model with the following state-of-the-art models: 
\begin{itemize}
    \item Nested Cache N-gram model \cite{hellendoorn2017deep}: an improved N-gram model which considers the unlimited vocabulary, nested scope, locality, and dynamism in source code.
    \item Pointer Mixture Network \cite{Li2018Code}: an attention and pointer-generator network-based code completion model.
\end{itemize}

The results are shown in Table \ref{tab:results}. \citet{hellendoorn2017deep} offers jar\footnote{https://github.com/SLP-team/SLP-Core/releases} to run their model. The input of their model is the token sequence, and the output is the accuracy of the next token's prediction on the whole dataset. In our datasets, the source code is represented as the AST node sequence. Each node has a type attribute, and the non-leaf nodes do not have a value attribute. We can only get the complete type sequence as input data for their model. So there are no results on the next value prediction. 

As can be seen from the results, on all the three datasets, our model outperforms all the baselines on both the next node's type and value prediction. For the next node's type prediction, our model achieves the accuracy of 87\%, 82\%, and 91\% on these three datasets respectively, which improves Nested N-gram model by 51\%, 40\%, and 72\%, and improves Pointer Mixture Network by 33\%, 24\%, and 24\%, in terms of \textit{normalized improvement in accuracy}. For the next node's value prediction, our model achieves the accuracy of 73\%, 73\%, and 83\% on three datasets, which improves Pointer Mixture Network by 16\%, 15\%, and 13\%, in terms of \textit{normalized improvement in accuracy}. In the value prediction, the predictions of the \textit{UNK} targets are treated as wrong predictions. The \textit{UNK} rates for Python, Java, and JavaScript are 11\%, 13\%, and 7\%. Therefore, when computing the \textit{normalized improvement in accuracy}, the upper bounds of the accuracy for the three datasets are 89\%, 87\%, and 93\%, not 100\%. In \citet{Li2018Code}'s model, Pointer Network is adopted to address the OoV issue in the value prediction. Different from their model, our model does not introduce the pointer network and can still outperform them. We apply the Wilcoxon Rank Sum Test (WRST) \cite{wilcoxon1945individual} to test whether the improvements of our model over baselines are statistically significant, and all the p-values are less than 1e-5, which indicates significant improvements. We also use Cliff's Delta \cite{macbeth2011cliff} to measure the effect size, and the values are non-negligible. From Table \ref{tab:results}, we also notice that the improvements on the JavaScript are not as good as the other two datasets. The reason might lie in that the correlation between the node’s type and value in JavaScript is weaker than Python and Java. As shown in Table \ref{tab:datasets}, the category of the node’s type for JavaScript is much less (only 95 types) compared with Python or Java, but one type can correspond to many values, which could result in the limited improvement.

\textbf{\textit{RQ2: What is the effectiveness of each component for our proposed model?}} ~ We perform an ablation study to examine the effects of two proposed components used in our model: the Multi-task Learning mechanism and the new path2root encoder. We conduct experiments without either MTL or path2root encoder, and we also conduct experiments on the vanilla Transformer-XL network by removing both of these two components. Besides, to verify whether capturing the long-range dependency from the input programs helps, we also conduct an experiment of removing the recurrent mechanism from the Transformer-XL architecture. The results are shown in Table \ref{tab:components}. The first row shows the results of our full model. The second row presents the results of removing MTL from the full model, and the third row removes the path2root encoder from the full model. The results of removing the recurrent mechanism from the Transformer-XL architecture are shown in the fourth row. The results of the vanilla Transformer-XL are shown in the last row. As seen from the results, removing either MTL or the path2root encoder results in a drop in the accuracy, and removing MTL drops more, which demonstrates that both the Multi-task Learning mechanism and the path2root encoder are necessary to improve the performance, and MTL contributes more to the improvements. When removing the recurrent mechanism from our full model, the accuracy drops a lot, even lower than the vanilla Transformer-XL network. These results demonstrate that capturing long-range dependency is of great importance and necessity for language modeling, and it serves as the basis of other improvements made in this paper. The statistical testing also shows that the improvements are statistically significant, and the effect sizes are non-negligible. 

\section{Discussion}\label{Discussion}
\subsection{Learning Process Analysis.}
\begin{figure}[t]
\setlength{\abovecaptionskip}{0cm}
\footnotesize
    \begin{flushleft}
    \begin{tikzpicture}
        \begin{axis}[width=0.75\columnwidth,
                     height=0.5\columnwidth,
                     xlabel=Epoch,
                     ylabel=Loss,
                     legend pos= outer north east,
                     legend style={font=\tiny},
                     ymajorgrids=true,
                     grid style=dashed]
\addplot[draw=blue,mark=diamond*] coordinates {(1,1.938)(2,1.712)(3,1.612)(4,1.543)(5,1.496)(6,1.465)(7,1.446)(8,1.434)};
\addlegendentry{Train: baseline}
\addplot[draw=blue,mark=square*] coordinates {(1,1.874)(2,1.827)(3,1.827)(4,1.848)(5,1.887)(6,1.923)(7,1.952)(8,1.964)};
\addlegendentry{Test: baseline}
\addplot[draw=red,mark=diamond*,color=red] coordinates {(1,1.81)(2,1.639)(3,1.58)(4,1.543)(5,1.52)(6,1.505)(7,1.496)(8,1.49)};
\addlegendentry{Train: our model}
\addplot[draw=red,mark=square*,color=red] coordinates {(1,1.72)(2,1.69)(3,1.68)(4,1.65)(5,1.63)(6,1.61)(7,1.59)(8,1.59)};
\addlegendentry{Test: our model}
        \end{axis}
    \end{tikzpicture}
    \end{flushleft}
    \caption{The cross-entropy loss on training and test set for baseline model and our model.}
    \label{Fig:training_process}
    \vspace{-0.3cm}
\end{figure}
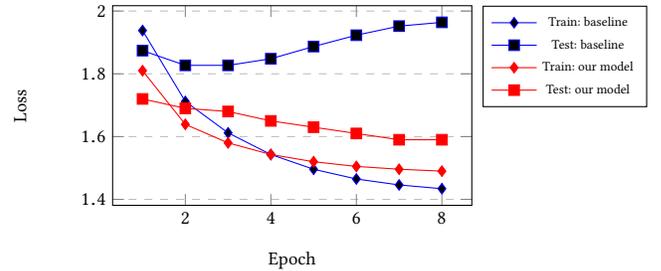

To find out why our proposed model performs better, we analyze the learning process of the state-of-the-art baseline model (Pointer Mixture Network \cite{Li2018Code}) and our proposed model. Figure \ref{Fig:training_process} shows the loss of predicting the next node's type after every epoch on Python's training and test set for the two models. As seen from the figure, the difference between the training loss and test loss is large in the baseline model, which is obviously the result of over-fitting. While in our model, the difference is much smaller. Furthermore, the test loss of our model is lower than the baseline model at each epoch. The reason lies in three aspects: (1) by utilizing the hierarchical structural information of AST and the information contained in the training signals of related tasks, our proposed model can extract more accurate and common features from programs, and thus can achieve better performance; (2) adopting the Transformer-XL architecture to model the long-range dependency in the programs helps our model capture more information from the context and thus improves model's performance; (3) multi-task learning provides an effective regularization method through knowledge sharing among tasks, thus can improve the model's performance by decreasing the difference between training and test loss, which to some extent prevents the model from over-fitting. For another two datasets, i.e., Java and JavaScript, we have the same observations and findings. 

\subsection{Training Cost Analysis}
To evaluate the cost of the improvements, we count the number of parameters and record the training time of our model and \citet{Li2018Code}'s model. To evaluate the cost of our proposed components, we also present these statistics data of the vanilla Transformer-XL network and removing one of the components from our model. Due to the page limitation, we take the training time in the Python dataset as an example. The run-time in the test process is very fast (about 0.1 milliseconds per query), and the difference in the test time among different models is little. Thus, we do not compare the test time. The number of trainable parameters and the training time are presented in Table \ref{tab:training_time}. 

\begin{table}[t]
  \begin{center}
  \setlength{\abovecaptionskip}{0cm} 
  \caption{Training cost analysis in the Python dataset.}
    \scalebox{0.9}{
    \begin{tabular}{lccc} 
     \toprule
      Model & \# of Parameters  & Training Time \\
      \midrule
      Pointer Mixture Network & 162.6M & 34 hours \\
      vanilla 6-layer Transformer-XL & 95.8M & 15 hours\\
      \midrule
      our model & 98.9M & 25 hours \\
      - MTL & 96.8M & 22 hours \\
      - Path2root Encoder & 97.6M & 20 hours \\ 
      \bottomrule
    \end{tabular}
    }
    \label{tab:training_time}
    \vspace{-0.3cm}
  \end{center}
\end{table}

For the number of training parameters, the 6-layer Transformer-XL network uses only 59\% of the parameter budget compared to Pointer Mixture Network \cite{Li2018Code} but can achieve comparable performance with them. In our model, we adopt Transformer-XL as the language model and apply Multi-task Learning to learn two tasks jointly and propose a new path2root encoder, which leads to an increase of the trainable parameters compared with the vanilla Transformer-XL networks. In our framework, the partial AST encoder, path2root encoder are shared among all tasks, and only the output layers are task-specific. Thus, the parameter increasing is slight, only by 3.2\% (from 95.8M to 98.9M). But the number of trainable parameters of our model is only 60.8\% of the number of trainable parameters in Pointer Mixture Network. Besides, we also count the number of the parameters of removing MTL or Path2root encoder from our model, and the results are presented in the last two rows in Table \ref{tab:training_time}. The results demonstrate that the additional parameters of integrating these two components into Transformer-XL increase a small number of parameters. 

For the training time, our full model spends 74\% of the time compared to Pointer Mixture Network \cite{Li2018Code}. In Pointer Mixture Network, they adopt LSTM as the language model, where most of the recurrent computations are performed during the hidden states' updating process. While in our model, Transformer-XL \cite{transformer-xl} is used as the language model. In Transformer-XL, the representations of each input for each segment are computed relying on the self-attention layers, and the recurrence only happens between segments. Thus, it allows for substantially more parallelization and requires less time to train. When removing MTL, the training time decreases slightly (from 25 hours to 22 hours) because most of the parameters are shared between tasks. Thus, applying MTL will not introduce much additional training time during the training process. Adding a path2root encoder into our model is an improvement towards the model's structure. It increases the model's complexity, which leads to increased training time. When removing the path2root encoder from our full model, the training time is reduced by 5 hours. Compared to vanilla Transformer-XL, applying the MTL and Path2root encoder will increase the training time, but considering the improvements, the increase is acceptable.

In summary, our model uses 59\% of the parameter budget and spends 74\% of the run-time to train compared to Pointer Mixture Network \cite{Li2018Code}, and can still outperform them statistically significant and by a substantial margin. We also have the same observations and results for the other two datasets, i.e., Java and JavaScript.   

\subsection{Effect of Weights for Task-specific Loss.} \label{weight}
In our MTL-based model, we use a weighted sum over task-specific losses as the final loss. By default, we set the weights for the two tasks as 0.5 and 0.5. The performance of the model is related to the choice of weighting between the tasks' loss. To show the effect of the weights, we present the results of different weight settings on our model in Table \ref{tab:weights}. $\alpha_1$ is the weight of the loss for the next node's type prediction task, and $\alpha_2$ is the weight of the loss for the next node's value prediction task. When one of the weights is set to 0, the model becomes a single-task model. As expected, when giving more weight to a task's loss, the accuracy of this task will be increased. However, when assigning a high weight to one task  (e.g., set $\alpha_1$ or $\alpha_2$ as 1), the advantage of the MTL would be affected, which results in poor performance. 

\begin{table}[t]
\setlength{\abovecaptionskip}{0cm} 
\caption{The results of different weight settings in our model.}
  \begin{center}
  \scalebox{0.9}{
    \begin{tabular}{llccccccc}
      \toprule
      \multirow{2}*{$\alpha_1$} & \multirow{2}*{$\alpha_2$} & \multicolumn{2}{c}{Python} & \multicolumn{2}{c}{Java} & \multicolumn{2}{c}{JavaScript} \\
      ~ & ~ & Type & Value & Type & Value & Type & Value \\
      \midrule
      \multicolumn{2}{c}{\citet{Li2018Code}} & 80.6\% & 70.1\% & 75.9\% & 70.7\% & 88.6\% & 81.0\% \\ 
      1.0 & 0 & 84.2\% & - & 79.7\% & - & 89.5\% & - \\
      0.7 & 0.3 & \bf86.9\% & 71.5\% & \textbf{81.7\%} & 71.6\% & \bf91.3\% & 80.3\% \\
      0.5 & 0.5 & 85.4\% & 72.0\% & 80.8\% & 72.7\% & 90.8\% & 81.0\% \\
      0.3 & 0.7 & 83.9\% & \textbf{73.2\%} & 79.8\%
      &\textbf{ 73.1\%} & 89.5\% & \textbf{82.5\%} \\
      0 & 1.0 & - & 71.8\% & - & 71.6\% & - & 80.1\%  \\
      \bottomrule
    \end{tabular}
  }
\label{tab:weights}
\vspace{-0.3cm}
\end{center}
\end{table}

\subsection{Qualitative Analysis}
\noindent\textbf{Difficult type predictions.} ~ Predicting the structure of the code, such as loops, if statements, and exception handling statements, is overall a very hard task \cite{raychev2016probabilistic}. \citet{raychev2016probabilistic} define a set of types on JavaScript that are hard to predict and name them as ``difficult type prediction''. We evaluate our model's performance on these types' prediction and compare our model with Pointer Mixture Network \cite{Li2018Code} on the same test set. The results are shown in Table \ref{tab:difficult_type}. As seen from the table, our model outperforms Pointer Mixture Network by a large margin in all these types. Besides, in our model, the variance of the accuracies for predicting each token is much smaller than the Pointer Mixture Network. The accuracies are mostly distributed in the range of 88\% - 93\%. In Pointer Mixture Network, the accuracies of those low-frequency tokens are very low. For example, ``SwitchStatement'' only appears 2625 times in the test set, the accuracy is only 45.9\% in Pointer Mixture Network. While in our model, the accuracy is 88.2\%, which is much higher than the Pointer Mixture Network. These results demonstrate that our model can discover the structure of programs and achieve an excellent generalization performance on structure predictions. 

\begin{table}[t] 
  \begin{center}
  \setlength{\abovecaptionskip}{0cm} 
  \setlength{\abovecaptionskip}{0cm} 
  \caption{Difficult type predictions on JavaScript}
    \begin{tabular}{lcc} 
      \toprule
      Difficult Type & Pointer Mixture Network & Our Model\\
      \midrule
      ContinueStatement & 65.6\% & \textbf{88.5\%}\\
      ForStatement &  65.5\% & \textbf{89.0\%} \\
      WhileStatement & 79.8\% & \textbf{88.9\%} \\
      ReturnStatement  & 61.4\% & \textbf{89.0\%} \\
      SwitchStatement & 45.9\% & \textbf{88.2\%} \\
      ThrowStatement & 54.1\% & \textbf{88.0\%} \\
      TryStatement  & 57.3\% & \textbf{88.9\%} \\
      IfStatement & 68.3\% &\textbf{89.0\% }\\
      \bottomrule
    \end{tabular}
    \label{tab:difficult_type}
    \vspace{-0.3cm}
  \end{center}
\end{table}

\begin{figure*}[h] 
\setlength{\abovecaptionskip}{0.2cm}
\centering\includegraphics[width=12.5cm,height=15cm]{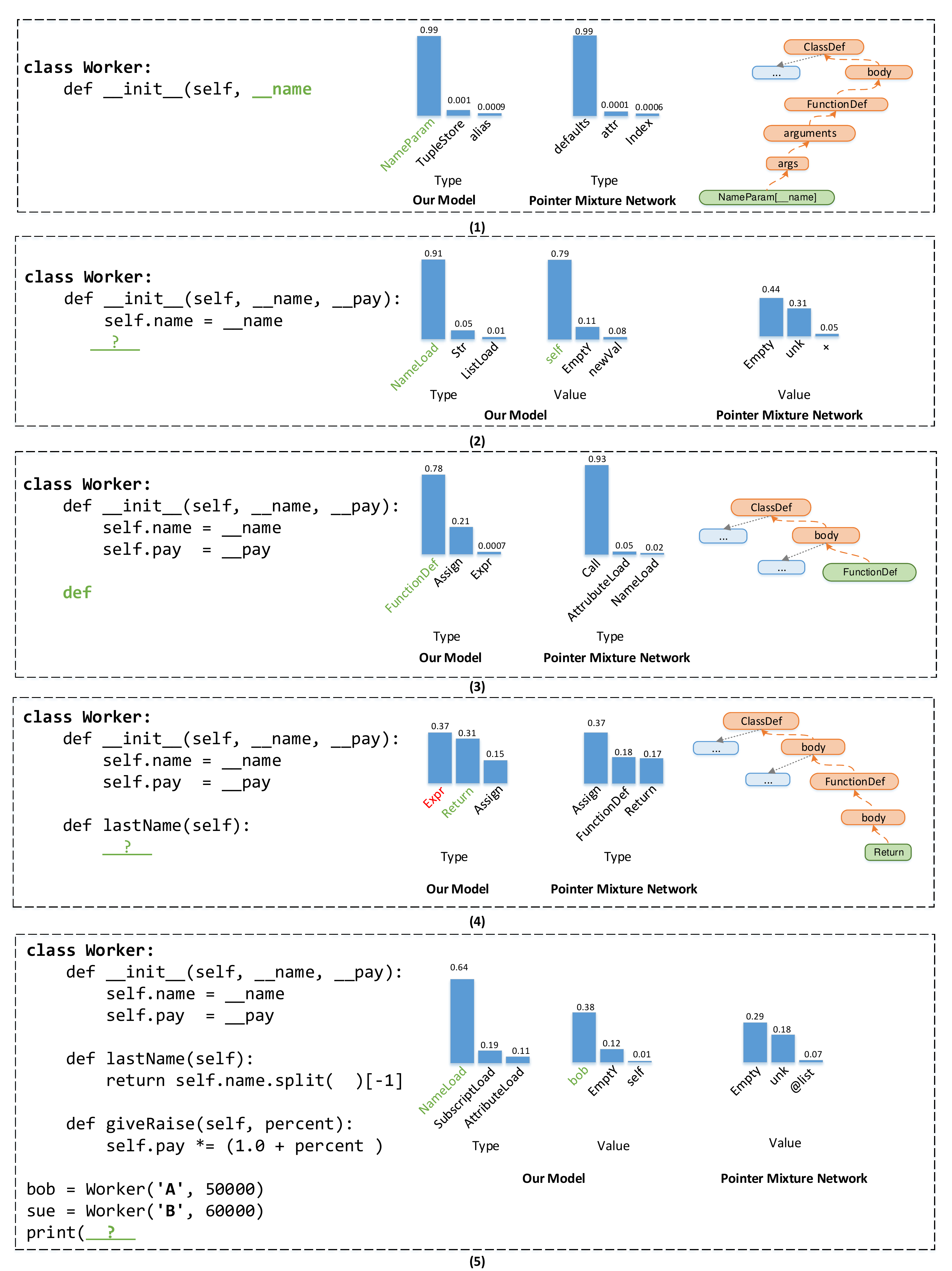}
\caption{Code completion examples.}
\vspace{-0.3cm}
\label{Fig:case}
\end{figure*}

\noindent\textbf{Example completion.} ~ 
Here, we present code completion examples on Python to analyze the performance of our proposed model. We take several positions in a Python code snippet to test the performance of our model and the baseline model. We show the top three predictions of our model and the baseline model of Pointer Mixture Network \cite{Li2018Code}. The results are shown in Figure \ref{Fig:case}. We divide the cases of the prediction into two situations:

\noindent a) \textbf{The effect of the path information}. In the first example, the target prediction \textit{\_\_name} is a parameter for the function \textit{\_\_init\_\_}, and its corresponding node's type is \textit{NameParam}. The path from it to the root node (shown on the right side of the example) implies the information that the prediction is a parameter of a function, thus it can help our model to make the correct prediction on the node's type. For the baseline model, it can only learn from the sequential context and fail to produce the right prediction. Similarly, in the third example, the target prediction \textit{def} means a function definition, where its corresponding node's type is \textit{FunctionDef}. With the information contained in the path, our model can make the correct prediction, while the baseline model fails. In the fourth example, both of our model and the baseline model fail to produce the correct prediction \textit{return}. In this case, the path cannot offer accurate information because there exist many possible children for a function's body. Thus, our model produces \textit{Expr}, which is also a grammatical child. The correct prediction is ranked second in our model and is ranked third in the baseline model. In cases like this, our model might make wrong predictions.

\noindent b) \textbf{The effect of MTL}. In the second example, the target prediction \textit{self} is not a new variable and has been used in the previous context. By correctly predicting \textit{NameLoad} in the node's type prediction task, our model can realize the value of the node is an already used value in the previous context. Thus it can identify the value from the context through the pointer. For the baseline model, it may not realize the prediction is a variable accessing operation without the help of the auxiliary task. Thus, it just predicts \textit{EMPTY} which is the most frequent node's value in our corpus. The last example is also in the same way.

\subsection{Threats to Validity}\label{limitation}
\noindent\textbf{Threats to external validity} relate to the quality of the datasets we used and the generalizability of our results. Python and JavaScript are two benchmarked datasets that have been used in previous code completion work \cite{raychev2016probabilistic,liu2016neural,Li2018Code}. Java dataset we used is from \citet{hu2018summarizing}. All of the programs in the datasets are collected from GitHub repositories, and each dataset contains 100,000 training programs and 50,000 test programs. However, further studies are needed to validate and generalize our findings to other programming languages. Furthermore, our case study is small scale. More user evaluation is needed to confirm and improve the usefulness of our code completion model.

\noindent\textbf{Threats to internal validity} include the influence of the weightings between each task's loss i.e., $\alpha_k$. The performance of our model would be affected by the different weights (discussed in Section \ref{weight}), which are tuned by hand in our experiments. However, the default weight settings of 0.5 and 0.5 for the next node's type and value prediction loss can still achieve a considerable performance increase. Take the experiments on the Python dataset as an example, default weight setting achieves 5\% (from 80.6\% to 85.4\%) improvements in accuracy on the next node's type prediction compared with \citet{Li2018Code}, which are only 1.5\% lower than the best weight settings. And the results in the next node's value prediction are also similar. Another threat to internal validity relates to the errors in the implementation of the baseline methods. For \citet{hellendoorn2017deep}, we directly used their published jars. Thus, there is little threat to approach implementation.

\noindent\textbf{Threats to construct validity} relate to the suitability of our evaluation measure. We use \textit{accuracy} as the metric which evaluates the proportion of correctly predicted next node's type or value. It is a classical evaluation measure for code completion and is used in almost all the previous code completion work \cite{hindle2012naturalness, tu2014localness, raychev2016probabilistic, hellendoorn2017deep, Li2018Code}.  

\section{Related Work}\label{related work}
\noindent\textbf{Code Completion} ~ 
Code completion is a hot research topic in the field of software engineering. Early work in code completion bases on on heuristic rules and static type information to make suggestions \cite{hou2010towards}, or bases on similar code examples \cite{bruch2009learning} and program history data \cite{robbes2008program}. Since \citet{hindle2012naturalness} found that source code contained predictable statistical properties, statistical language models began to be used for modeling source code \cite{nguyen2013statistical,tu2014localness,hellendoorn2017deep,Li2018Code}, where N-gram is the most widely used model. \cite{tu2014localness} observed that source code has a unique property of localness, which could not be captured by the traditional N-gram model. They improved N-gram by adding a cache mechanism to exploit localness and achieved better performance than other N-gram based models. \citet{hellendoorn2017deep} introduced an improved N-gram model that considered the unlimited vocabulary, nested scope, locality, and dynamism in source code. Their evaluation results on code completion showed that their model outperformed existing statistical language models, including deep learning based models. Thus we choose their model as a baseline. \citet{raychev2016probabilistic} proposed a probabilistic model based on decision tree and domain-specific grammars. They performed experiments to predict AST nodes on Python and JavaScript datasets.  

In recent years, deep recurrent neural network-based language models have been applied to learning source code and have made great progress \cite{white2015toward,bhoopchand2016learning,Li2018Code}. \citet{liu2016neural} proposed a code completion model based on a vanilla LSTM network. \citet{bhoopchand2016learning} proposed an RNN model with a sparse pointer mechanism aiming at capturing long-range dependencies. \citet{Li2018Code} proposed a pointer mixture network to address the OoV issue. For the next node's type prediction, their model outperforms \citet{raychev2016probabilistic} on both Python and JavaScript datasets. For the next node's value prediction, their model outperforms \citet{raychev2016probabilistic} on Python and achieves comparable performance on JavaScript. \citet{Li2018Code} has achieved state-of-the-art results, which is used as a baseline in this paper. In the above work, RNNs, in particular, LSTM neural network-based language models are adopted to model the programs. However, these techniques are found not sufficient to model the long-term dependencies in the sequential data \cite{khandelwal2018sharp}. In our model, we adopt Transformer-XL \cite{transformer-xl} as the language model to capture the long-range dependencies in the programs. Besides, we also propose a novel method to introduce the hierarchical structural information into the program's representation, which is not well considered in previous code completion work. 

\noindent\textbf{Multi-task Learning} ~ Multi-task learning has been used successfully across many fields including natural language processing \cite{liu2015representation,guo2018soft,devlin2018bert}, speech recognition \cite{deng2013new} and computer vision \cite{long2015learning,lu2017fully}. In the natural language processing area, MTL has been proven effectively in many tasks, such as machine translation \cite{luong2015multi,dong2015multi,zaremoodi2018adaptive}, text summarization \cite{isonuma2017extractive,guo2018soft}, and sequence labeling \cite{peng2016multi,lin2018multi}. However, to the best of our knowledge, MTL has not been applied to programming language processing yet. In code completion, there exist several related tasks. For example, predicting the next node's type and value in AST. Existing code completion models perform a specific task in one model, which leads to the underuse of information from related tasks. In this paper, we apply MTL to code completion to predict the next node's type and value jointly and improve the state-of-the-art statistically significant and substantially.

\section{Conclusions and Future Work}\label{conclusion}
In this paper, we propose an MTL-based self-attentional neural architecture for code completion. For code representation, we propose a novel method to model the hierarchical information of the predicting node explicitly. To capture the long-term dependency in the programs, we apply the Transformer-XL network as the base language model. For the model's learning process, we apply MTL to enable knowledge sharing between related tasks. Experimental results demonstrate that the proposed model achieves better results than previous state-of-the-art models. To the best of our knowledge, we are the first to apply MTL and Transformer-XL to code completion. We believe this work represents a significant advance in programming language modeling, which will be beneficial as a building block for many other applications in this area. 

In the future, we plan to improve the effectiveness of our proposed model by introducing domain-specific constraints such as grammar rules.

\begin{acks}
This research is supported by the National Key R\&D Program under Grant No. 2018YFB1003904, the National Natural Science Foundation of China under Grant No. 61832009, No. 61620106007 and No. 61751210, and the Australian Research Council's Discovery Early Career Researcher Award (DECRA) funding scheme (DE200100021).
\end{acks}

\bibliographystyle{ACM-Reference-Format}
\bibliography{sample-base}


\begin{thebibliography}{40}

%%% the Web address.

\ifx \showCODEN    \undefined \def \showCODEN     #1{\unskip}     \fi
\ifx \showDOI      \undefined \def \showDOI       #1{#1}\fi
\ifx \showISBNx    \undefined \def \showISBNx     #1{\unskip}     \fi
\ifx \showISBNxiii \undefined \def \showISBNxiii  #1{\unskip}     \fi
\ifx \showISSN     \undefined \def \showISSN      #1{\unskip}     \fi
\ifx \showLCCN     \undefined \def \showLCCN      #1{\unskip}     \fi
\ifx \shownote     \undefined \def \shownote      #1{#1}          \fi
\ifx \showarticletitle \undefined \def \showarticletitle #1{#1}   \fi
\ifx \showURL      \undefined \def \showURL       {\relax}        \fi
% invisible to TeX
\providecommand\bibfield[2]{#2}
\providecommand\bibinfo[2]{#2}
\providecommand\natexlab[1]{#1}
\providecommand\showeprint[2][]{arXiv:#2}

\bibitem[\protect\citeauthoryear{Abadi, Barham, Chen, Chen, Davis, Dean, Devin,
  Ghemawat, Irving, Isard, et~al\mbox{.}}{Abadi et~al\mbox{.}}{2016}]%
        {abadi2016tensorflow}
\bibfield{author}{\bibinfo{person}{Mart{\'\i}n Abadi}, \bibinfo{person}{Paul
  Barham}, \bibinfo{person}{Jianmin Chen}, \bibinfo{person}{Zhifeng Chen},
  \bibinfo{person}{Andy Davis}, \bibinfo{person}{Jeffrey Dean},
  \bibinfo{person}{Matthieu Devin}, \bibinfo{person}{Sanjay Ghemawat},
  \bibinfo{person}{Geoffrey Irving}, \bibinfo{person}{Michael Isard},
  {et~al\mbox{.}}} \bibinfo{year}{2016}\natexlab{}.
\newblock \showarticletitle{Tensorflow: A system for large-scale machine
  learning}. In \bibinfo{booktitle}{\emph{12th $\{$USENIX$\}$ Symposium on
  Operating Systems Design and Implementation ($\{$OSDI$\}$ 16)}}.
  \bibinfo{pages}{265--283}.
\newblock


\bibitem[\protect\citeauthoryear{Bahdanau, Cho, and Bengio}{Bahdanau
  et~al\mbox{.}}{2015}]%
        {bahdanau2014neural}
\bibfield{author}{\bibinfo{person}{Dzmitry Bahdanau},
  \bibinfo{person}{Kyunghyun Cho}, {and} \bibinfo{person}{Yoshua Bengio}.}
  \bibinfo{year}{2015}\natexlab{}.
\newblock \showarticletitle{Neural Machine Translation by Jointly Learning to
  Align and Translate}.
\newblock  (\bibinfo{year}{2015}).
\newblock


\bibitem[\protect\citeauthoryear{Bhoopchand, Rockt{\"{a}}schel, Barr, and
  Riedel}{Bhoopchand et~al\mbox{.}}{2016}]%
        {bhoopchand2016learning}
\bibfield{author}{\bibinfo{person}{Avishkar Bhoopchand}, \bibinfo{person}{Tim
  Rockt{\"{a}}schel}, \bibinfo{person}{Earl~T. Barr}, {and}
  \bibinfo{person}{Sebastian Riedel}.} \bibinfo{year}{2016}\natexlab{}.
\newblock \showarticletitle{Learning Python Code Suggestion with a Sparse
  Pointer Network}.
\newblock \bibinfo{journal}{\emph{CoRR}}  \bibinfo{volume}{abs/1611.08307}
  (\bibinfo{year}{2016}).
\newblock
\showeprint[arxiv]{1611.08307}
\urldef\tempurl%
\url{http://arxiv.org/abs/1611.08307}
\showURL{%
\tempurl}


\bibitem[\protect\citeauthoryear{Bruch, Monperrus, and Mezini}{Bruch
  et~al\mbox{.}}{2009}]%
        {bruch2009learning}
\bibfield{author}{\bibinfo{person}{Marcel Bruch}, \bibinfo{person}{Martin
  Monperrus}, {and} \bibinfo{person}{Mira Mezini}.}
  \bibinfo{year}{2009}\natexlab{}.
\newblock \showarticletitle{Learning from examples to improve code completion
  systems}. In \bibinfo{booktitle}{\emph{Proceedings of the 7th joint meeting
  of the European software engineering conference and the ACM SIGSOFT symposium
  on the foundations of software engineering}}. \bibinfo{pages}{213--222}.
\newblock


\bibitem[\protect\citeauthoryear{Caruana}{Caruana}{1997}]%
        {caruana1997multitask}
\bibfield{author}{\bibinfo{person}{Rich Caruana}.}
  \bibinfo{year}{1997}\natexlab{}.
\newblock \showarticletitle{Multitask Learning}.
\newblock \bibinfo{journal}{\emph{Machine Learning}} \bibinfo{volume}{28},
  \bibinfo{number}{1} (\bibinfo{year}{1997}), \bibinfo{pages}{41--75}.
\newblock


\bibitem[\protect\citeauthoryear{Cho, van Merrienboer, Bahdanau, and
  Bengio}{Cho et~al\mbox{.}}{2014}]%
        {cho2014properties}
\bibfield{author}{\bibinfo{person}{Kyunghyun Cho}, \bibinfo{person}{Bart van
  Merrienboer}, \bibinfo{person}{Dzmitry Bahdanau}, {and}
  \bibinfo{person}{Yoshua Bengio}.} \bibinfo{year}{2014}\natexlab{}.
\newblock \showarticletitle{On the Properties of Neural Machine Translation:
  Encoder-Decoder Approaches}.
\newblock  (\bibinfo{year}{2014}), \bibinfo{pages}{103--111}.
\newblock


\bibitem[\protect\citeauthoryear{Costa, Figueiredo, Murta, and Sarma}{Costa
  et~al\mbox{.}}{2016}]%
        {costa2016tipmerge}
\bibfield{author}{\bibinfo{person}{Catarina Costa}, \bibinfo{person}{Jair
  Figueiredo}, \bibinfo{person}{Leonardo Murta}, {and} \bibinfo{person}{Anita
  Sarma}.} \bibinfo{year}{2016}\natexlab{}.
\newblock \showarticletitle{TIPMerge: recommending experts for integrating
  changes across branches}. In \bibinfo{booktitle}{\emph{Proceedings of the
  2016 24th ACM SIGSOFT International Symposium on Foundations of Software
  Engineering}}. ACM, \bibinfo{pages}{523--534}.
\newblock


\bibitem[\protect\citeauthoryear{Dai, Yang, Yang, Carbonell, Le, and
  Salakhutdinov}{Dai et~al\mbox{.}}{2019}]%
        {transformer-xl}
\bibfield{author}{\bibinfo{person}{Zihang Dai}, \bibinfo{person}{Zhilin Yang},
  \bibinfo{person}{Yiming Yang}, \bibinfo{person}{Jaime~G. Carbonell},
  \bibinfo{person}{Quoc~Viet Le}, {and} \bibinfo{person}{Ruslan
  Salakhutdinov}.} \bibinfo{year}{2019}\natexlab{}.
\newblock \showarticletitle{Transformer-XL: Attentive Language Models beyond a
  Fixed-Length Context}. In \bibinfo{booktitle}{\emph{Proceedings of the 57th
  Conference of the Association for Computational Linguistics, {ACL} 2019,
  Florence, Italy, July 28- August 2, 2019, Volume 1: Long Papers}}.
  \bibinfo{pages}{2978--2988}.
\newblock


\bibitem[\protect\citeauthoryear{Deng, Hinton, and Kingsbury}{Deng
  et~al\mbox{.}}{2013}]%
        {deng2013new}
\bibfield{author}{\bibinfo{person}{Li Deng}, \bibinfo{person}{Geoffrey~E.
  Hinton}, {and} \bibinfo{person}{Brian Kingsbury}.}
  \bibinfo{year}{2013}\natexlab{}.
\newblock \showarticletitle{New types of deep neural network learning for
  speech recognition and related applications: an overview}. In
  \bibinfo{booktitle}{\emph{{IEEE} International Conference on Acoustics,
  Speech and Signal Processing, {ICASSP} 2013, Vancouver, BC, Canada, May
  26-31, 2013}}. \bibinfo{publisher}{{IEEE}}, \bibinfo{pages}{8599--8603}.
\newblock


\bibitem[\protect\citeauthoryear{Devlin, Chang, Lee, and Toutanova}{Devlin
  et~al\mbox{.}}{2018}]%
        {devlin2018bert}
\bibfield{author}{\bibinfo{person}{Jacob Devlin}, \bibinfo{person}{Ming{-}Wei
  Chang}, \bibinfo{person}{Kenton Lee}, {and} \bibinfo{person}{Kristina
  Toutanova}.} \bibinfo{year}{2018}\natexlab{}.
\newblock \showarticletitle{{BERT:} Pre-training of Deep Bidirectional
  Transformers for Language Understanding}.
\newblock \bibinfo{journal}{\emph{CoRR}}  \bibinfo{volume}{abs/1810.04805}
  (\bibinfo{year}{2018}).
\newblock
\showeprint[arxiv]{1810.04805}
\urldef\tempurl%
\url{http://arxiv.org/abs/1810.04805}
\showURL{%
\tempurl}


\bibitem[\protect\citeauthoryear{Dong, Wu, He, Yu, and Wang}{Dong
  et~al\mbox{.}}{2015}]%
        {dong2015multi}
\bibfield{author}{\bibinfo{person}{Daxiang Dong}, \bibinfo{person}{Hua Wu},
  \bibinfo{person}{Wei He}, \bibinfo{person}{Dianhai Yu}, {and}
  \bibinfo{person}{Haifeng Wang}.} \bibinfo{year}{2015}\natexlab{}.
\newblock \showarticletitle{Multi-Task Learning for Multiple Language
  Translation}. In \bibinfo{booktitle}{\emph{Proceedings of the 53rd Annual
  Meeting of the Association for Computational Linguistics and the 7th
  International Joint Conference on Natural Language Processing of the Asian
  Federation of Natural Language Processing, {ACL} 2015, July 26-31, 2015,
  Beijing, China, Volume 1: Long Papers}}. \bibinfo{publisher}{The Association
  for Computer Linguistics}, \bibinfo{pages}{1723--1732}.
\newblock


\bibitem[\protect\citeauthoryear{Guo, Pasunuru, and Bansal}{Guo
  et~al\mbox{.}}{2018}]%
        {guo2018soft}
\bibfield{author}{\bibinfo{person}{Han Guo}, \bibinfo{person}{Ramakanth
  Pasunuru}, {and} \bibinfo{person}{Mohit Bansal}.}
  \bibinfo{year}{2018}\natexlab{}.
\newblock \showarticletitle{Soft Layer-Specific Multi-Task Summarization with
  Entailment and Question Generation}. In \bibinfo{booktitle}{\emph{Proceedings
  of the 56th Annual Meeting of the Association for Computational Linguistics,
  {ACL} 2018, Melbourne, Australia, July 15-20, 2018, Volume 1: Long Papers}}.
  \bibinfo{publisher}{Association for Computational Linguistics},
  \bibinfo{pages}{687--697}.
\newblock


\bibitem[\protect\citeauthoryear{Hellendoorn and Devanbu}{Hellendoorn and
  Devanbu}{2017}]%
        {hellendoorn2017deep}
\bibfield{author}{\bibinfo{person}{Vincent~J. Hellendoorn} {and}
  \bibinfo{person}{Premkumar~T. Devanbu}.} \bibinfo{year}{2017}\natexlab{}.
\newblock \showarticletitle{Are deep neural networks the best choice for
  modeling source code?}. In \bibinfo{booktitle}{\emph{Proceedings of the 2017
  11th Joint Meeting on Foundations of Software Engineering, {ESEC/FSE} 2017,
  Paderborn, Germany, September 4-8, 2017}}. \bibinfo{publisher}{{ACM}},
  \bibinfo{pages}{763--773}.
\newblock


\bibitem[\protect\citeauthoryear{Hindle, Barr, Su, Gabel, and Devanbu}{Hindle
  et~al\mbox{.}}{2012}]%
        {hindle2012naturalness}
\bibfield{author}{\bibinfo{person}{Abram Hindle}, \bibinfo{person}{Earl~T.
  Barr}, \bibinfo{person}{Zhendong Su}, \bibinfo{person}{Mark Gabel}, {and}
  \bibinfo{person}{Premkumar~T. Devanbu}.} \bibinfo{year}{2012}\natexlab{}.
\newblock \showarticletitle{On the naturalness of software}. In
  \bibinfo{booktitle}{\emph{34th International Conference on Software
  Engineering, {ICSE} 2012, June 2-9, 2012, Zurich, Switzerland}}.
  \bibinfo{publisher}{{IEEE} Computer Society}, \bibinfo{pages}{837--847}.
\newblock


\bibitem[\protect\citeauthoryear{Hochreiter and Schmidhuber}{Hochreiter and
  Schmidhuber}{1997}]%
        {hochreiter1997long}
\bibfield{author}{\bibinfo{person}{Sepp Hochreiter} {and}
  \bibinfo{person}{J{\"{u}}rgen Schmidhuber}.} \bibinfo{year}{1997}\natexlab{}.
\newblock \showarticletitle{Long Short-Term Memory}.
\newblock \bibinfo{journal}{\emph{Neural Computation}} \bibinfo{volume}{9},
  \bibinfo{number}{8} (\bibinfo{year}{1997}), \bibinfo{pages}{1735--1780}.
\newblock


\bibitem[\protect\citeauthoryear{Hou and Pletcher}{Hou and Pletcher}{2010}]%
        {hou2010towards}
\bibfield{author}{\bibinfo{person}{Daqing Hou} {and} \bibinfo{person}{David~M
  Pletcher}.} \bibinfo{year}{2010}\natexlab{}.
\newblock \showarticletitle{Towards a better code completion system by API
  grouping, filtering, and popularity-based ranking}. In
  \bibinfo{booktitle}{\emph{Proceedings of the 2nd International Workshop on
  Recommendation Systems for Software Engineering}}. \bibinfo{pages}{26--30}.
\newblock


\bibitem[\protect\citeauthoryear{Hu, Li, Xia, Lo, Lu, and Jin}{Hu
  et~al\mbox{.}}{2018}]%
        {hu2018summarizing}
\bibfield{author}{\bibinfo{person}{Xing Hu}, \bibinfo{person}{Ge Li},
  \bibinfo{person}{Xin Xia}, \bibinfo{person}{David Lo}, \bibinfo{person}{Shuai
  Lu}, {and} \bibinfo{person}{Zhi Jin}.} \bibinfo{year}{2018}\natexlab{}.
\newblock \showarticletitle{Summarizing Source Code with Transferred {API}
  Knowledge}. In \bibinfo{booktitle}{\emph{Proceedings of the Twenty-Seventh
  International Joint Conference on Artificial Intelligence, {IJCAI} 2018, July
  13-19, 2018, Stockholm, Sweden.}} \bibinfo{publisher}{ijcai.org},
  \bibinfo{pages}{2269--2275}.
\newblock


\bibitem[\protect\citeauthoryear{Isonuma, Fujino, Mori, Matsuo, and
  Sakata}{Isonuma et~al\mbox{.}}{2017}]%
        {isonuma2017extractive}
\bibfield{author}{\bibinfo{person}{Masaru Isonuma}, \bibinfo{person}{Toru
  Fujino}, \bibinfo{person}{Junichiro Mori}, \bibinfo{person}{Yutaka Matsuo},
  {and} \bibinfo{person}{Ichiro Sakata}.} \bibinfo{year}{2017}\natexlab{}.
\newblock \showarticletitle{Extractive Summarization Using Multi-Task Learning
  with Document Classification}. In \bibinfo{booktitle}{\emph{Proceedings of
  the 2017 Conference on Empirical Methods in Natural Language Processing,
  {EMNLP} 2017, Copenhagen, Denmark, September 9-11, 2017}}.
  \bibinfo{publisher}{Association for Computational Linguistics},
  \bibinfo{pages}{2101--2110}.
\newblock


\bibitem[\protect\citeauthoryear{Khandelwal, He, Qi, and Jurafsky}{Khandelwal
  et~al\mbox{.}}{2018}]%
        {khandelwal2018sharp}
\bibfield{author}{\bibinfo{person}{Urvashi Khandelwal}, \bibinfo{person}{He
  He}, \bibinfo{person}{Peng Qi}, {and} \bibinfo{person}{Dan Jurafsky}.}
  \bibinfo{year}{2018}\natexlab{}.
\newblock \showarticletitle{Sharp Nearby, Fuzzy Far Away: How Neural Language
  Models Use Context}.
\newblock  (\bibinfo{year}{2018}), \bibinfo{pages}{284--294}.
\newblock


\bibitem[\protect\citeauthoryear{Kingma and Ba}{Kingma and Ba}{2015}]%
        {KingmaB14}
\bibfield{author}{\bibinfo{person}{Diederik~P. Kingma} {and}
  \bibinfo{person}{Jimmy Ba}.} \bibinfo{year}{2015}\natexlab{}.
\newblock \showarticletitle{Adam: {A} Method for Stochastic Optimization}. In
  \bibinfo{booktitle}{\emph{3rd International Conference on Learning
  Representations, {ICLR} 2015, San Diego, CA, USA, May 7-9, 2015, Conference
  Track Proceedings}}, \bibfield{editor}{\bibinfo{person}{Yoshua Bengio} {and}
  \bibinfo{person}{Yann LeCun}} (Eds.).
\newblock


\bibitem[\protect\citeauthoryear{Li, Wang, Lyu, and King}{Li
  et~al\mbox{.}}{2018}]%
        {Li2018Code}
\bibfield{author}{\bibinfo{person}{Jian Li}, \bibinfo{person}{Yue Wang},
  \bibinfo{person}{Michael~R. Lyu}, {and} \bibinfo{person}{Irwin King}.}
  \bibinfo{year}{2018}\natexlab{}.
\newblock \showarticletitle{Code Completion with Neural Attention and Pointer
  Networks}. In \bibinfo{booktitle}{\emph{Proceedings of the Twenty-Seventh
  International Joint Conference on Artificial Intelligence, {IJCAI} 2018, July
  13-19, 2018, Stockholm, Sweden.}} \bibinfo{publisher}{ijcai.org},
  \bibinfo{pages}{4159--4165}.
\newblock
\urldef\tempurl%
\url{https://doi.org/10.24963/ijcai.2018/578}
\showURL{%
\tempurl}


\bibitem[\protect\citeauthoryear{Lin, Yang, Stoyanov, and Ji}{Lin
  et~al\mbox{.}}{2018}]%
        {lin2018multi}
\bibfield{author}{\bibinfo{person}{Ying Lin}, \bibinfo{person}{Shengqi Yang},
  \bibinfo{person}{Veselin Stoyanov}, {and} \bibinfo{person}{Heng Ji}.}
  \bibinfo{year}{2018}\natexlab{}.
\newblock \showarticletitle{A Multi-lingual Multi-task Architecture for
  Low-resource Sequence Labeling}. In \bibinfo{booktitle}{\emph{Proceedings of
  the 56th Annual Meeting of the Association for Computational Linguistics,
  {ACL} 2018, Melbourne, Australia, July 15-20, 2018, Volume 1: Long Papers}}.
  \bibinfo{publisher}{Association for Computational Linguistics},
  \bibinfo{pages}{799--809}.
\newblock


\bibitem[\protect\citeauthoryear{Liu, Wang, Shin, Gonzalez, and Song}{Liu
  et~al\mbox{.}}{2016}]%
        {liu2016neural}
\bibfield{author}{\bibinfo{person}{Chang Liu}, \bibinfo{person}{Xin Wang},
  \bibinfo{person}{Richard Shin}, \bibinfo{person}{Joseph~E Gonzalez}, {and}
  \bibinfo{person}{Dawn Song}.} \bibinfo{year}{2016}\natexlab{}.
\newblock \showarticletitle{Neural Code Completion}.
\newblock  (\bibinfo{year}{2016}).
\newblock


\bibitem[\protect\citeauthoryear{Liu, Gao, He, Deng, Duh, and Wang}{Liu
  et~al\mbox{.}}{2015}]%
        {liu2015representation}
\bibfield{author}{\bibinfo{person}{Xiaodong Liu}, \bibinfo{person}{Jianfeng
  Gao}, \bibinfo{person}{Xiaodong He}, \bibinfo{person}{Li Deng},
  \bibinfo{person}{Kevin Duh}, {and} \bibinfo{person}{Ye{-}Yi Wang}.}
  \bibinfo{year}{2015}\natexlab{}.
\newblock \showarticletitle{Representation Learning Using Multi-Task Deep
  Neural Networks for Semantic Classification and Information Retrieval}. In
  \bibinfo{booktitle}{\emph{{NAACL} {HLT} 2015, The 2015 Conference of the
  North American Chapter of the Association for Computational Linguistics:
  Human Language Technologies, Denver, Colorado, USA, May 31 - June 5, 2015}}.
  \bibinfo{publisher}{The Association for Computational Linguistics},
  \bibinfo{pages}{912--921}.
\newblock


\bibitem[\protect\citeauthoryear{Long and Wang}{Long and Wang}{2015}]%
        {long2015learning}
\bibfield{author}{\bibinfo{person}{Mingsheng Long} {and}
  \bibinfo{person}{Jianmin Wang}.} \bibinfo{year}{2015}\natexlab{}.
\newblock \showarticletitle{Learning Multiple Tasks with Deep Relationship
  Networks}.
\newblock \bibinfo{journal}{\emph{CoRR}}  \bibinfo{volume}{abs/1506.02117}
  (\bibinfo{year}{2015}).
\newblock
\showeprint[arxiv]{1506.02117}
\urldef\tempurl%
\url{http://arxiv.org/abs/1506.02117}
\showURL{%
\tempurl}


\bibitem[\protect\citeauthoryear{Lu, Kumar, Zhai, Cheng, Javidi, and Feris}{Lu
  et~al\mbox{.}}{2017}]%
        {lu2017fully}
\bibfield{author}{\bibinfo{person}{Yongxi Lu}, \bibinfo{person}{Abhishek
  Kumar}, \bibinfo{person}{Shuangfei Zhai}, \bibinfo{person}{Yu Cheng},
  \bibinfo{person}{Tara Javidi}, {and} \bibinfo{person}{Rog{\'{e}}rio~Schmidt
  Feris}.} \bibinfo{year}{2017}\natexlab{}.
\newblock \showarticletitle{Fully-Adaptive Feature Sharing in Multi-Task
  Networks with Applications in Person Attribute Classification}. In
  \bibinfo{booktitle}{\emph{2017 {IEEE} Conference on Computer Vision and
  Pattern Recognition, {CVPR} 2017, Honolulu, HI, USA, July 21-26, 2017}}.
  \bibinfo{publisher}{{IEEE} Computer Society}, \bibinfo{pages}{1131--1140}.
\newblock


\bibitem[\protect\citeauthoryear{Luong, Le, Sutskever, Vinyals, and
  Kaiser}{Luong et~al\mbox{.}}{2016}]%
        {luong2015multi}
\bibfield{author}{\bibinfo{person}{Minh{-}Thang Luong},
  \bibinfo{person}{Quoc~V. Le}, \bibinfo{person}{Ilya Sutskever},
  \bibinfo{person}{Oriol Vinyals}, {and} \bibinfo{person}{Lukasz Kaiser}.}
  \bibinfo{year}{2016}\natexlab{}.
\newblock \showarticletitle{Multi-task Sequence to Sequence Learning}.
\newblock  (\bibinfo{year}{2016}).
\newblock


\bibitem[\protect\citeauthoryear{Macbeth, Razumiejczyk, and Ledesma}{Macbeth
  et~al\mbox{.}}{2011}]%
        {macbeth2011cliff}
\bibfield{author}{\bibinfo{person}{Guillermo Macbeth}, \bibinfo{person}{Eugenia
  Razumiejczyk}, {and} \bibinfo{person}{Rub{\'e}n~Daniel Ledesma}.}
  \bibinfo{year}{2011}\natexlab{}.
\newblock \showarticletitle{Cliff's Delta Calculator: A non-parametric effect
  size program for two groups of observations}.
\newblock \bibinfo{journal}{\emph{Universitas Psychologica}}
  \bibinfo{volume}{10}, \bibinfo{number}{2} (\bibinfo{year}{2011}),
  \bibinfo{pages}{545--555}.
\newblock


\bibitem[\protect\citeauthoryear{Mou, Li, Zhang, Wang, and Jin}{Mou
  et~al\mbox{.}}{2016}]%
        {mou2016convolutional}
\bibfield{author}{\bibinfo{person}{Lili Mou}, \bibinfo{person}{Ge Li},
  \bibinfo{person}{Lu Zhang}, \bibinfo{person}{Tao Wang}, {and}
  \bibinfo{person}{Zhi Jin}.} \bibinfo{year}{2016}\natexlab{}.
\newblock \showarticletitle{Convolutional Neural Networks over Tree Structures
  for Programming Language Processing}. In
  \bibinfo{booktitle}{\emph{Proceedings of the Thirtieth {AAAI} Conference on
  Artificial Intelligence, February 12-17, 2016, Phoenix, Arizona, {USA.}}}
  \bibinfo{publisher}{{AAAI} Press}, \bibinfo{pages}{1287--1293}.
\newblock


\bibitem[\protect\citeauthoryear{Nguyen, Nguyen, Nguyen, and Nguyen}{Nguyen
  et~al\mbox{.}}{2013}]%
        {nguyen2013statistical}
\bibfield{author}{\bibinfo{person}{Tung~Thanh Nguyen},
  \bibinfo{person}{Anh~Tuan Nguyen}, \bibinfo{person}{Hoan~Anh Nguyen}, {and}
  \bibinfo{person}{Tien~N. Nguyen}.} \bibinfo{year}{2013}\natexlab{}.
\newblock \showarticletitle{A statistical semantic language model for source
  code}. In \bibinfo{booktitle}{\emph{Joint Meeting of the European Software
  Engineering Conference and the {ACM} {SIGSOFT} Symposium on the Foundations
  of Software Engineering, ESEC/FSE'13, Saint Petersburg, Russian Federation,
  August 18-26, 2013}}. \bibinfo{publisher}{{ACM}}, \bibinfo{pages}{532--542}.
\newblock


\bibitem[\protect\citeauthoryear{Peng and Dredze}{Peng and Dredze}{2017}]%
        {peng2016multi}
\bibfield{author}{\bibinfo{person}{Nanyun Peng} {and} \bibinfo{person}{Mark
  Dredze}.} \bibinfo{year}{2017}\natexlab{}.
\newblock \showarticletitle{Multi-task Domain Adaptation for Sequence Tagging}.
  In \bibinfo{booktitle}{\emph{Proceedings of the 2nd Workshop on
  Representation Learning for NLP, Rep4NLP@ACL 2017, Vancouver, Canada, August
  3, 2017}}. \bibinfo{publisher}{Association for Computational Linguistics},
  \bibinfo{pages}{91--100}.
\newblock


\bibitem[\protect\citeauthoryear{Raychev, Bielik, and Vechev}{Raychev
  et~al\mbox{.}}{2016}]%
        {raychev2016probabilistic}
\bibfield{author}{\bibinfo{person}{Veselin Raychev}, \bibinfo{person}{Pavol
  Bielik}, {and} \bibinfo{person}{Martin~T. Vechev}.}
  \bibinfo{year}{2016}\natexlab{}.
\newblock \showarticletitle{Probabilistic model for code with decision trees}.
  In \bibinfo{booktitle}{\emph{Proceedings of the 2016 {ACM} {SIGPLAN}
  International Conference on Object-Oriented Programming, Systems, Languages,
  and Applications, {OOPSLA} 2016, part of {SPLASH} 2016, Amsterdam, The
  Netherlands, October 30 - November 4, 2016}}. \bibinfo{publisher}{{ACM}},
  \bibinfo{pages}{731--747}.
\newblock


\bibitem[\protect\citeauthoryear{Robbes and Lanza}{Robbes and Lanza}{2008}]%
        {robbes2008program}
\bibfield{author}{\bibinfo{person}{Romain Robbes} {and}
  \bibinfo{person}{Michele Lanza}.} \bibinfo{year}{2008}\natexlab{}.
\newblock \showarticletitle{How program history can improve code completion}.
  In \bibinfo{booktitle}{\emph{2008 23rd IEEE/ACM International Conference on
  Automated Software Engineering}}. IEEE, \bibinfo{pages}{317--326}.
\newblock


\bibitem[\protect\citeauthoryear{Ruder}{Ruder}{2017}]%
        {ruder2017overview}
\bibfield{author}{\bibinfo{person}{Sebastian Ruder}.}
  \bibinfo{year}{2017}\natexlab{}.
\newblock \showarticletitle{An Overview of Multi-Task Learning in Deep Neural
  Networks}.
\newblock \bibinfo{journal}{\emph{CoRR}}  \bibinfo{volume}{abs/1706.05098}
  (\bibinfo{year}{2017}).
\newblock
\showeprint[arxiv]{1706.05098}
\urldef\tempurl%
\url{http://arxiv.org/abs/1706.05098}
\showURL{%
\tempurl}


\bibitem[\protect\citeauthoryear{Schuster and Paliwal}{Schuster and
  Paliwal}{1997}]%
        {schuster1997bidirectional}
\bibfield{author}{\bibinfo{person}{Mike Schuster} {and}
  \bibinfo{person}{Kuldip~K. Paliwal}.} \bibinfo{year}{1997}\natexlab{}.
\newblock \showarticletitle{Bidirectional recurrent neural networks}.
\newblock \bibinfo{journal}{\emph{{IEEE} Trans. Signal Processing}}
  \bibinfo{volume}{45}, \bibinfo{number}{11} (\bibinfo{year}{1997}),
  \bibinfo{pages}{2673--2681}.
\newblock


\bibitem[\protect\citeauthoryear{Tu, Su, and Devanbu}{Tu et~al\mbox{.}}{2014}]%
        {tu2014localness}
\bibfield{author}{\bibinfo{person}{Zhaopeng Tu}, \bibinfo{person}{Zhendong Su},
  {and} \bibinfo{person}{Premkumar~T. Devanbu}.}
  \bibinfo{year}{2014}\natexlab{}.
\newblock \showarticletitle{On the localness of software}. In
  \bibinfo{booktitle}{\emph{Proceedings of the 22nd {ACM} {SIGSOFT}
  International Symposium on Foundations of Software Engineering, (FSE-22),
  Hong Kong, China, November 16 - 22, 2014}}. \bibinfo{publisher}{{ACM}},
  \bibinfo{pages}{269--280}.
\newblock


\bibitem[\protect\citeauthoryear{Vaswani, Shazeer, Parmar, Uszkoreit, Jones,
  Gomez, Kaiser, and Polosukhin}{Vaswani et~al\mbox{.}}{2017}]%
        {vaswani2017attention}
\bibfield{author}{\bibinfo{person}{Ashish Vaswani}, \bibinfo{person}{Noam
  Shazeer}, \bibinfo{person}{Niki Parmar}, \bibinfo{person}{Jakob Uszkoreit},
  \bibinfo{person}{Llion Jones}, \bibinfo{person}{Aidan~N Gomez},
  \bibinfo{person}{{\L}ukasz Kaiser}, {and} \bibinfo{person}{Illia
  Polosukhin}.} \bibinfo{year}{2017}\natexlab{}.
\newblock \showarticletitle{Attention is all you need}. In
  \bibinfo{booktitle}{\emph{Advances in neural information processing
  systems}}. \bibinfo{pages}{5998--6008}.
\newblock


\bibitem[\protect\citeauthoryear{White, Vendome, V{\'{a}}squez, and
  Poshyvanyk}{White et~al\mbox{.}}{2015}]%
        {white2015toward}
\bibfield{author}{\bibinfo{person}{Martin White}, \bibinfo{person}{Christopher
  Vendome}, \bibinfo{person}{Mario~Linares V{\'{a}}squez}, {and}
  \bibinfo{person}{Denys Poshyvanyk}.} \bibinfo{year}{2015}\natexlab{}.
\newblock \showarticletitle{Toward Deep Learning Software Repositories}. In
  \bibinfo{booktitle}{\emph{12th {IEEE/ACM} Working Conference on Mining
  Software Repositories, {MSR} 2015, Florence, Italy, May 16-17, 2015}}.
  \bibinfo{publisher}{{IEEE} Computer Society}, \bibinfo{pages}{334--345}.
\newblock


\bibitem[\protect\citeauthoryear{Wilcoxon}{Wilcoxon}{1945}]%
        {wilcoxon1945individual}
\bibfield{author}{\bibinfo{person}{Frank Wilcoxon}.}
  \bibinfo{year}{1945}\natexlab{}.
\newblock \showarticletitle{Individual comparisons by ranking methods}.
\newblock \bibinfo{journal}{\emph{Biometrics bulletin}} \bibinfo{volume}{1},
  \bibinfo{number}{6} (\bibinfo{year}{1945}), \bibinfo{pages}{80--83}.
\newblock


\bibitem[\protect\citeauthoryear{Zaremoodi, Buntine, and Haffari}{Zaremoodi
  et~al\mbox{.}}{2018}]%
        {zaremoodi2018adaptive}
\bibfield{author}{\bibinfo{person}{Poorya Zaremoodi}, \bibinfo{person}{Wray~L.
  Buntine}, {and} \bibinfo{person}{Gholamreza Haffari}.}
  \bibinfo{year}{2018}\natexlab{}.
\newblock \showarticletitle{Adaptive Knowledge Sharing in Multi-Task Learning:
  Improving Low-Resource Neural Machine Translation}. In
  \bibinfo{booktitle}{\emph{Proceedings of the 56th Annual Meeting of the
  Association for Computational Linguistics, {ACL} 2018, Melbourne, Australia,
  July 15-20, 2018, Volume 2: Short Papers}}. \bibinfo{publisher}{Association
  for Computational Linguistics}, \bibinfo{pages}{656--661}.
\newblock


\end{thebibliography}


\end{document}